\documentclass[3p]{elsarticle}

\usepackage{epsfig}
\usepackage{amsfonts}
\usepackage{amssymb,amsmath}
\usepackage{graphics}
\usepackage{mathrsfs}
\usepackage{setspace} 

\begin{document}
	
\begin{frontmatter}	
	
\title{\bf How one can assess  the chaos?}

\author{Marat Akhmet$^{*,a}$} \ead{marat@metu.edu.tr}

\address{$^a$Department of Mathematics, Middle East Technical University, 06800 Ankara, Turkey}

\address{$^*$Corresponding Author Tel.:+90 541  718 9447}

\begin{abstract}
We propose an uncertainty principle for chaos, focusing on two key characteristics: alpha unpredictability and Lorenz sensitivity. This principle outlines a limitation on the relationship between two infinite sequences that underpin these concepts. It is applicable to both deterministic and stochastic dynamics, marking a significant step toward integrating these two fields. Our initial progress in this area was achieved through research on Markov chains utilizing alpha labeling.

Additionally, we offer suggestions on how this principle can assess the degree  of chaos in specific processes. We also outline open questions regarding the relationships among various types of chaos, including a modification of the recurrence theorem.
\end{abstract}

\end{frontmatter}

\section{Chaos and the principle as a unity of opposites}

Heraclitus was the first to articulate the law of the unity of opposites. Opposites exist within our minds and serve as tools for analysis. This law is connected to N. Bohr's principle of complementarity and W. Heisenberg's uncertainty principle. Consequently, it is useful to view opposites as related to chaos. The most common manifestation of this unity is found in the concepts of divergence and convergence. In various types of chaos, these concepts are evident in Lorenz sensitivity \cite{Lorenz, Rossler, Chua, Yorke}. Ultra Poincar\'e chaos represents a new phenomenon in complexity \cite{Akhmet1, Akhmet2, Akhmet3}, characterized by alpha unpredictability and Poisson stability,  which pertain to  convergence and divergence respectively.

Any form of uncertainty is a result of opposition. Drawing on ideas from conventionalism \cite{Ben}, one could argue that opposites arise from the observation of uncertainty. Therefore, it is important to recognize that position and momentum are opposites, as described by Heisenberg's law. It is not surprising that the uncertainty principle related to chaos, which is suggested for discussion, is grounded in these obvious opposites of convergence and divergence.

In our research on chaos, we have identified uncertainty in dynamical chaos by employing the following inequalities:
\begin{equation}\label{eq1}
	h_{\alpha} \le s_n d(p_n,p)\le h_{\beta},   
\end{equation}
where \( h_{\alpha} \) and \( h_{\beta} \) are positive numbers, \( s_n \) represents the time sequence of divergence, \( p_n \) denotes the converging space sequence, and \( d \) is a metric. This formula has been presented in the book \cite{Akhmet1}. We believe that these inequalities will provide insights into chaos that are akin to those found in quantum mechanics. Related topics concerning alpha unpredictability, ultra Poincaré and Lorenz chaos, artificial neural networks \cite{Akhmet4}, differential equations, deterministic and stochastic dynamical systems \cite{Akhmet1}, and synchronization \cite{Akhmet1} can be explored through this proposal.

%Open problems related to the theory of functions, the recurrence theorem, and chaos investigation will be announced. 

The concept of the alpha unpredictable point serves as the foundation for ultra Poincaré chaos \cite{Akhmet1}. A point $p$ in a dynamic system is considered alpha unpredictable if there exists a positive number $e$ and sequences $ t_n$ and $s_n$ that both increase toward infinity. In this context, let $p_n = f(t_n, p)$ converge to  $p$, while the points $f(s_n, p_n)$ maintain a distance greater than $e$ from  $f(s_n, p)$ for all $n$.

The uncertainty inherent in alpha unpredictability arises from the inability to determine the infinite sequence of pairs $(s_n, p_n)$  for $n = 1, 2, \ldots$ Consequently, chaos cannot be complitely observed in experiments and simulations. As we observe closer approximation points $f(t_n, p)$, the moments of divergence $s_n$ become larger. Conversely, as we fine-tune a large moment $s_n$, identifying the corresponding $p_n$ becomes increasingly challenging. A more precisely determined $s_n$ results in less accuracy for $p_n$. Therefore, we grapple with this uncertainty, and the sequences of points $p_n$ and moments $s_n$ are just as theoretical as  position and momentum of Heisenberg's uncertainty principle.
Thus, we use the term  "uncertainty" to imply that chaos, under our conditions, can only be established if complete information about the sequences $t_n, s_n,$  and $p_n$ is collected. More specifically, we view $(s_n, p_n), n = 1, 2, \ldots$ as a chaotic pair. This discussion relates to alpha unpredictability in the context of ultra Poincaré chaos and the sensitivity characteristic of Lorenz chaos. Hence, we are addressing the principle of uncertainty of chaos as a phenomenon. In the following sections, we will elaborate on why this principle applies to all types of chaos. If we accept that we are dealing with the uncertainty of chaos concerning the sequences, it can be recognized that the inequalities in (\ref{eq1})  establish a principle of uncertainty since they provide limits (estimates) for the phenomenon, which is encompassed within numerical laws. Algorithms to calculate these sequences have already been developed in \cite{Akhmet1,Akhmet4,Akhmet5}. 
The sensitivity is a source of uncertainty in Lorenz chaos, which is the main ingredient of the sophisticated dynamics.   A motion  $f(q,t)$ is sensitive at a   point \textit{p} provided that there exists a positive number \textit{e},     sequences of points $p_n$  and moments $s_n, n =1,2,\ldots,$ such that $p_n \to \textit{p},$ and distance between $f(p,s_n)$  and $f(p_n,s_n)$ is larger than \textit{e} for all $n.$   
We have to look for points $p_n$ arbitrarily close to point \textit{p} to learn sensitivity,   but this makes less observable moments where the divergence happens.   Moreover, if one tries to determine large moments of the divergence,  then precision for the location of corresponding points $p_n$ decreases, and vice versa.  Finally, the infinite couples $(s_n,p_n), n = 1,2,\ldots$ can not be indicated ever. Thus, the principle of the uncertainty for chaos is based on Lorenz's sensitivity.
The separation number \textit{e}  can be an uncertainty constant for ultra Poincar\'e and Lorenz chaos since the smallness of error for the initial point does not affect the separation size.

The discussions above suggest that deterministic chaos, similar to quantum mechanics, is related to the  deep nature of reality. Furthermore, we believe that, based on the principle of chaos, one can identify analogs of wave functions and the Schrödinger equation for deterministic phenomena. We are confident that there are additional relationships between spatial and temporal sequences that should be acknowledged within the paradigm of uncertainty.

What is the philosophical conclusion of our findings? The greater the uncertainties in dynamics, the more valuable they become. Uncertainty is a key factor that indicates a theoretical phenomenon is worth investigating, as it embodies the unity of opposites. Therefore, discovering the uncertainty inherent in chaos is a significant achievement.

\section{More arguments for the principle} 

Based on our findings, we assert that ultra Poincaré chaos represents the most advanced form of irregular dynamics in the field of science. While other types of chaos rely on dense closed orbits as a foundational framework, we focus on unclosed Poisson stable orbits, each of which is dense within the chaotic region. This implies that the alternatives to ultra Poincaré chaos do not depend on the principle to the same extent that alpha unpredictability does. This could potentially explain the uncertainty observed in our studies.

Next, we will address the necessity of the positive constants, $h_{\alpha}$ and $h_{\beta}$. One might inquire whether chaos can exist without these constraints. Below, we present two arguments to address this question.

(a) Assume that a constant $h_{\beta}$ for chaotic dynamics cannot be identified as a finite number. In this case, there is a possibility that a pair $(s_n, p_n)$, for $n = 1, 2, \ldots$, satisfies $s_n \to \infty$, while the distance $d(p_n, p)$ remains bounded away from zero. This scenario indicates that the pair cannot be chaotic. This situation highlights that the existence of the constant is necessary to rule out the absence of chaos. The final remark about the constant is that the larger 
$h_{\beta}$ is, the closer the dynamics are to being non-chaotic and regular.

(b) Now, consider the case where there is no appropriate lower boundary, $h_{\alpha}$. Then  can find  as  a chaotic pair $(s_n, p_n)$,  where  $s_n$  is  a bounded sequence  and $d(p_n, p)$ approaches zero. This contradicts the principle of continuous dependence on initial values in dynamical systems. It is clear that the closer $h_{\alpha}$ is to zero in (\ref{eq1}), the more we advance toward quantum chaos.

Therefore, we assert that for chaos to exist, the relationships expressed in equations (\ref{eq1}) must hold true. More specifically, this means that one should be able to extract a subsequence from a chaotic pair that satisfies the principle. In other words, for a dynamical system $f(t, p)$ to be considered chaotic, it is essential that there exist sequences $s_n \to \infty$ and $p_n \to p$ such that the principle in equation (\ref{eq1}) is upheld.  The discussion on subsequences implies that one can say about  statistical nature of the principle.  A violation of the uncertainty undermines  guarantees of chaos.

Let us add more uncertainty to the dynamics by implication of the principle.

a) From the formula (\ref{eq1})  it implies that 

\begin{equation}\label{eq2}
	r^n_{\alpha} =\frac{h_{\alpha}}{s_n} \le  d(p_n,p)\le \frac{h_{\beta}}{s_n} = r^n_{\beta}.   
\end{equation}

The relations (\ref{eq2}) imply  that  corresponding to the  divergence moments $s_n$  points $p_n$ are in the strip of the width $r^n_{\beta} - r^n_{\alpha} = \frac{h_{\beta}- h_{\alpha} }{s_n}.$ The width tends to zero as $n \to \infty.$ Thus, the points are not out of the strip, not in the disc with radius  $r^n_{\alpha}$ and not out of the disc with radius    $r^n_{\beta}.$ Since the radiuses' smallness, this increases the uncertainty. 

b) Similarly, we have that 

\begin{equation}\label{eq3}
	R^n_{\alpha} =\frac{h_{\alpha}}{d(p_n,p)} \le  s_n \le \frac{h_{\beta}}{d(p_n,p)} = R^n_{\beta}.   
\end{equation}

The last formula shows that for $p_n \to p$ sequence $s_n$  is placed in the strip, whose width increases infinitely, but there are two intervals,  which should be free of the moments.  It is obvious that  the strip  existence also increases the uncertainty and chaos.

\section{How to apply  the principle to measure degree of chaos}

The uncertainty principle has not been previously discussed in the literature, even though various types of chaos such as  Devaney chaos, and Li-Yorke chaos have been examined for a considerable time. The reason for this oversight seems to be the periodic motions, which are a key component in the definitions of these chaotic systems. The presence of infinitely many closed orbits  in bounded domains supports the validity of inequalities (\ref{eq1}) for these systems, and they always present there implicitly.
We propose that proving the uncertainty principle should be regarded as an open problem for these types of chaos. Additionally, another open problem is to establish a connection between the Feigenbaum constant, which is derived from the period-doubling route to chaos, and the calculation of the uncertainty constant. In the context of ultra Poincaré chaos, we do not observe any prerequisites for the uncertainty principle, suggesting it may be a fundamental property of chaos.

Finally, we believe that under specific conditions, the constants associated with these chaotic systems may correlate with those defined by M. Planck and E. Hubble.

Now consider more general aspects related to (\ref{eq1}), specifically estimations suitable for all types of complex dynamics.

(a) An increase in $ h_{\beta} $ indicates that the divergence moments $ s_n$ are tending to infinity at a much faster rate than the points $ p_n $  approach the issue point $ p $. This suggests that the degree of irregularity is decreasing, meaning that chaos is diminishing. Geometrically, this could imply that the widths of the bands in the simulations of the Rössler and Chua models, which enclose the trajectories, are becoming wider.

%\begin{figure}[ht] 
%	\centerline{\includegraphics[width=7.5cm]{Definitions/gamma1}}
%	\vspace*{8pt}
%	\caption{The graph of the piecewise constant argument function $\gamma(t).$}
%	\label{gamma}
%\end{figure}

\begin{figure} [ht]  \label{Chua}
	\centering
	\includegraphics[width=5cm]{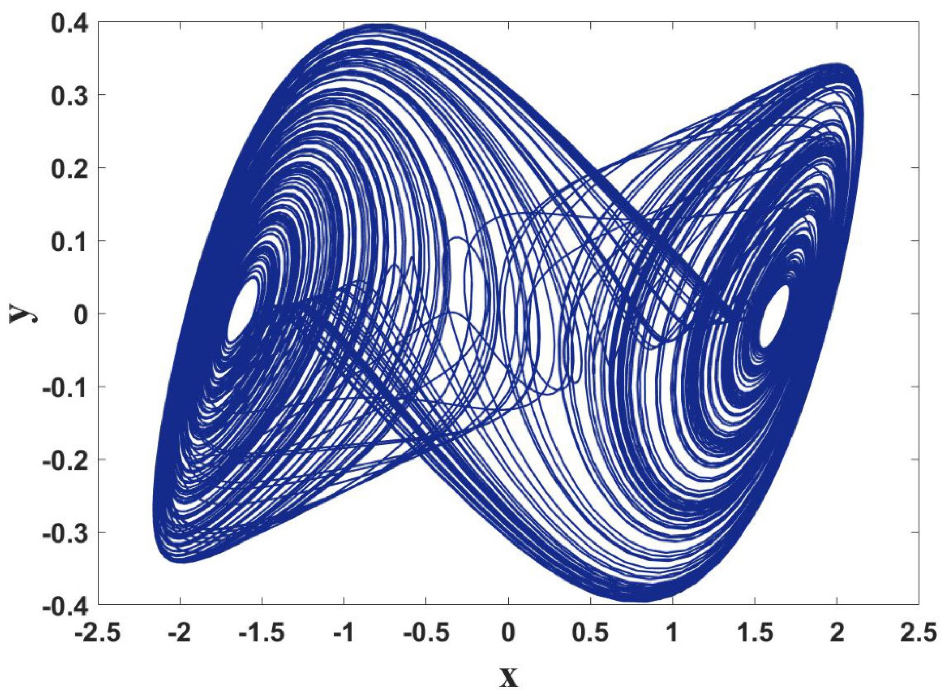}
	\vspace{20pt}
	\caption{Chua's attractor has specific features that ensure the existence of the fundamental constants.}
\end{figure}

(b) As the value of $ h_{\alpha}$ approaches zero, it indicates that the continuous dependence on the initial conditions in the dynamics becomes weaker. Ultimately, this can lead to stochastic processes, such as Bernoulli schemes or Markov chains, which we have previously studied, where the dependence on initial conditions disappears.

(c) To achieve a greater degree of qualitative chaos in a dynamical system, the difference between $ h_{\beta} $ and $ h_{\alpha}$ must  not be too large, and $ h_{\alpha}$  should be separated from zero. We can examine the Chua attractor and the Rössler band in this context. By analyzing the visual representations of these sets, we can infer that the chaos present in the attractor is stronger than that in the band. To validate our hypothesis, further simulations are needed.

\begin{figure}[ht] 
	\centering
	\includegraphics[height=5cm]{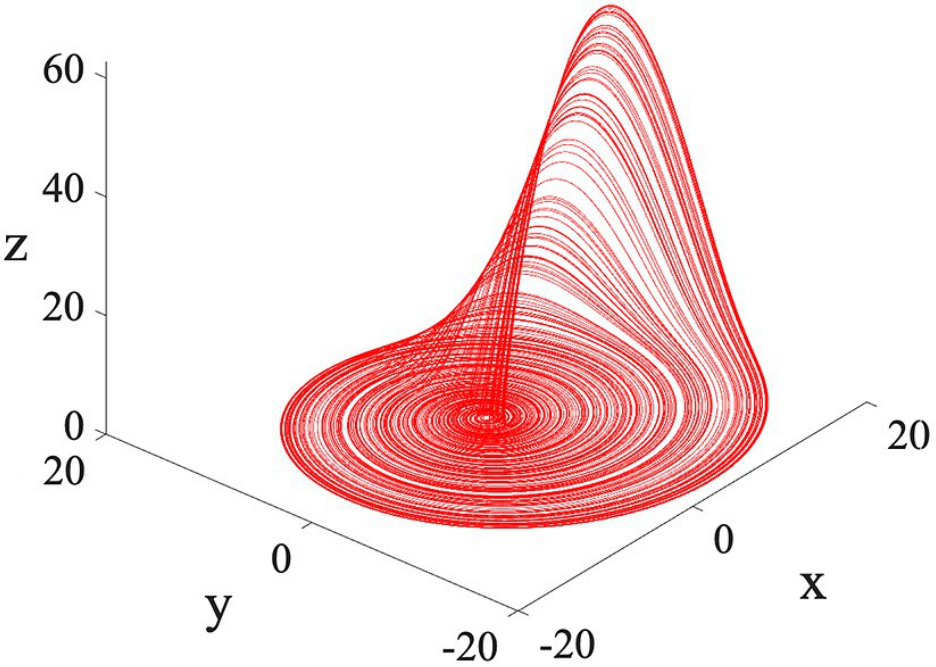}
	\vspace{20pt}
	\caption{ Rossler's band has qualitative characteristics that are sufficient to derive the principles of the attractor.}
	\label{Rossler}
\end{figure}

\section{What are the consequences for science?}

It is clear that the mentioned relations can be valuable for the qualitative theory of chaos and its manifestation in other fields of science. Potential issues are discussed below.

It is crucial to emphasize that the recurrence theorem \cite{Carat} specifically addresses Poisson stability and is regarded as a fundamental result in ergodicity. A key assumption of the theorem is the preservation of volume. We believe that this does not contradict the concept of alpha unpredictability. However, we do not think alpha unpredictability can be derived solely from this assumption. Therefore, the recurrence theorem can be adapted to incorporate alpha unpredictability and the uncertainty inherent in chaos and ergodicity.

We assert that this principle is applicable to any form of chaos exhibiting Lorenz sensitivity. This includes Li-Yorke and Devaney chaos, as well as those described by period-doubling bifurcation diagrams. Consequently, the principle serves as a necessary condition. It is of significant interest to confirm the presence of uncertainty based on specific characteristics such as dense closed orbits and bifurcation diagrams. We believe that the dynamical reasons underlying the Feigenbaum constant can be useful for the uncertainty constants outlined.

Alpha labeling is a new mathematical structure developed in our research to study complexities \cite{Akhmet1}. In our papers, this structure has been applied to chaos, fractals, and random processes. We believe that uncertainty constants can be determined for processes that are reduced to alpha labeling.

We hope that the principle of uncertainty within the dynamics of alpha labeling will be applicable to understanding uncertainty in stochastic processes with finite state spaces, such as Markov chains, and that it can later extend to continuous random processes. Additional applications are anticipated in the realm of quantum chaos, where relative restrictions stem from M. Planck's constant. If the qualitative characteristics introduced in our studies are applied to deterministic, stochastic, and quantum processes, it could represent a formal unification of scientific disciplines.

A key problem is to clarify the selection of uncertainty constants. Are they unique to each chaotic dynamic, or can they be derived from various fields such as chemistry, biology, or physics? How do these constants relate to quantum or cosmological constants? Is it possible to calculate them based on the Feigenbaum constant when considering the period-doubling route to chaos?

It is crucial to understand that there is no Bayesian probability within this principle since there is no uncertainty in the probability itself. Nevertheless, our study considers stochastic motions that include deterministic chaotic elements. This pertains to the sequences of convergence and divergence that we introduce with respect to alpha unpredictability.

To better reflect realistic dynamics, we propose using scenarios that can be developed from abstractions in Conway's Game of Life \cite{Con} and the dynamics of multiple quantum trajectories as analyzed by Feynman \cite{Fey}, particularly in relation to chaos unpredictability.

In conclusion, we express strong confidence that, based on alpha unpredictability and alpha labeling \cite{Akhmet1}, the principle of uncertainty in relation to chaos can significantly contribute to research on the complexities of the world.


\begin{thebibliography}{30}
	
	\bibitem{Lorenz} Lorenz, E.N., {\em Three approaches to atmospheric predictability,} Bulletin of the American Meteorological Society, {\bf 50(5)} (1969), 345-349.	
	
	
	\bibitem{Rossler} Rössler, O. E.,  {\em An Equation for Continuous Chaos,} Physics Letters, {\bf 57A(5)} (1976), 397–398.
	
	
	\bibitem{Chua} Matsumoto, T., {\em A Chaotic Attractor from Chua's Circuit,} IEEE Transactions on Circuits and System, {\bf CAS-31 (12)} (1984), 1055–1058.
	
	
	\bibitem{Yorke} Li,  T.-Y., Yorke, J.A., {\em Period Three Implies Chaos,} The American Mathematical Monthly, {\bf 82(10)} (1975), 985-992.
	
	\bibitem{Akhmet1}  Akhmet, M., {\em Ultra Poincaré Chaos and Alpha Labeling: A new approach to chaotic dynamics,} IOP 2024.
	
	\bibitem{Akhmet2} Akhmet, M.U., Fen, M.O., {\em Unpredictable points and chaos,} Communications in Nonlinear Science and Numerical Simulation {\bf 40} (2016), 1-5.
	
	\bibitem{Akhmet3} Akhmet, M.U., Fen, M.O.,{\em  Poincare chaos and unpredictable functions,}  Communications in Nonlinear Science and Numerical Simulation, {\bf 41} (2017), 85-94.	
	
	\bibitem {Ben}   Ben-Menahem,  Y., {\em  Conventionalism: From Poincar\'e  to Quine,}  Cambridge University Press 2006. 
	
	
	\bibitem{Akhmet4} Akhmet, M., Tleubergenova, M., Zhamanshin, A., Nugayeva, Z.  {\em Artificial Neural Networks: Alpha unpredictability and chaotic dynamics,} Springer 2025. 
	
	\bibitem{Akhmet5}  Akhmet, M.,  Baskan, K., Yesil, C.,   	 {\em Delta synchronization of Poincar\'e chaos in gas discharge-semiconductor systems}, Chaos,  {\bf 32(81)}  (2022), 083137.
	
	\bibitem{Carat}  Carath\'eodory, C.,  {\em   $\ddot{U}$ber den Wiederkehrsatz von Poincar\'e,}    Berl. Sitzungsber (1919), 580–584.
	
	\bibitem{Con}  Conway, J.H., {\em On numbers and games,} Academic Press 1976. 
	
	\bibitem{Fey}  Feynman, R.,  {\em The Theory of Positrons,} Physical Review, {\bf 76 (6)} 	(1949),  749–759.
	
\end{thebibliography}
\end{document}